\documentclass[journal]{IEEEtran}
\ifCLASSINFOpdf
\else
\fi
\usepackage{cite}
\usepackage{multicol}
\usepackage{amsmath,amssymb,amsfonts}
\usepackage{amsthm}
\usepackage{bm}
\usepackage{empheq}%

\usepackage{breqn}

\usepackage{graphicx}
\graphicspath{{Images/}}

\usepackage{booktabs}
\usepackage{multirow}
\usepackage{tabularx,multirow,ragged2e}
\newcolumntype{Y}{>{\RaggedRight\arraybackslash}X}
\newcolumntype{P}[1]{>{\RaggedRight\arraybackslash}p{#1}}

\usepackage[utf8]{inputenc}
\usepackage{dirtytalk}
\usepackage{tabu}

\usepackage{color}
\usepackage{comment}
\usepackage{xspace}
\usepackage{booktabs}
\usepackage{flushend}
\usepackage{float}

\usepackage{csquotes}

\usepackage{caption,subcaption}

\usepackage{algorithm}
\usepackage{algpseudocode}
\usepackage{algcompatible}
\usepackage{pifont}

\usepackage[dvipsnames]{xcolor}

\usepackage{url}

\usepackage{nomencl}
\makenomenclature

\usepackage{ifthen}
  \renewcommand{\nomgroup}[1]{%
  \item[\bfseries
  \ifthenelse{\equal{#1}{P}}{Sets, Parameters}{%
  \ifthenelse{\equal{#1}{V}}{Variables}{}}%
  ]}

\allowdisplaybreaks

\usepackage{nomencl}
\usepackage{xcolor}
\makenomenclature
 
\usepackage{etoolbox}
\renewcommand\nomgroup[1]{%
  \item[\bfseries
  \ifstrequal{#1}{S}{\textit{\textbf{Sets}}}{%
  \ifstrequal{#1}{V}{\textit{\textbf{Variables}}}{%
  \ifstrequal{#1}{I}{\textit{\textbf{Indices, Superscripts, Subscripts and Sets}}}{%
  \ifstrequal{#1}{P}{\textit{\textbf{Parameters}}}{}}}}%
]}

\usepackage{etoolbox}
\renewcommand\nomgroup[1]{%
  \item[\bfseries
  \ifstrequal{#1}{A}{\textit{\textbf{Indices, Superscripts, Subscripts and Sets: d-OPF}}}{%
  \ifstrequal{#1}{B}{\textit{\textbf{Indices, Superscripts, Subscripts and Sets: OTF}}}{%
  \ifstrequal{#1}{C}{\textit{\textbf{Indices, Superscripts, Subscripts and Sets: Algorithm}}}{%
  \ifstrequal{#1}{X}{\textit{\textbf{Variables: d-OPF}}}{%
  \ifstrequal{#1}{Y}{\textit{\textbf{Variables: OTF}}}{%
  \ifstrequal{#1}{Z}{\textit{\textbf{Variables: Algorithm}}}{%
  \ifstrequal{#1}{L}{\textit{\textbf{Parameters: d-OPF}}}{%
  \ifstrequal{#1}{M}{\textit{\textbf{Parameters: OTF}}}{%
  \ifstrequal{#1}{N}{\textit{\textbf{Parameters: Algorithm}}}{}}}}}}}}}%
]}

\DeclareMathOperator{\argmin}{arg\,min}

\algnewcommand\algorithmicor{\textbf{or}}

\newcommand*\rot{\rotatebox{90}}

\begin{document}
\title{A Novel Decentralized Algorithm for Coordinating the Optimal Power and Traffic Flows with EVs based on Variable Inner Loop Selection}
\author{Santosh~Sharma,~\IEEEmembership{Student~Member,~IEEE,}
       and~Qifeng~Li,~\IEEEmembership{Senior Member,~IEEE,} 
\thanks{Authors are with the Department of Electrical Engineering, University of Central Florida.}
\vspace{-8mm}}

\maketitle 
\begin{abstract}
The electric power distribution network (PDN) and the transportation network (TN) are generally operated/coordinated by different entities. However, they are coupled with each other due to electric vehicle charging stations (EVCSs). This paper proposes to coordinate the operation of the two systems via a fully decentralized framework where the PDN and TN operators solve their own operation problems by sharing only limited information. Nevertheless, the operation problems generally are in mixed-integer programming (MIP) form. To the best of our knowledge, the most existing decentralized/distributed optimization algorithms, such as the alternating direction method of multipliers (ADMM), are not always guaranteed to converge for such MIP problems. Therefore, a novel fully decentralized optimization algorithm is proposed, whose contributions include: 1) it is applicable to MIP problems with convergence and optimality guaranteed for mild assumptions, 2) it only requires limited information exchange between PDN and TN operators, which will help preserve the privacy of the two systems and reduce the investment in building communication channels, and 3) it is fully decentralized so that all the computations are carried out by PDN operator and TN coordinator only. Simulations on the test cases show the efficiency and efficacy of the proposed framework and algorithm.
\end{abstract}

\IEEEpeerreviewmaketitle
\vspace{-3mm}
\section{Introduction}
\IEEEPARstart{D}ue to the increasing concern over carbon emissions, electric vehicles (EVs) are gradually replacing fossil-fueled vehicles \cite{859fb3f78f3243d1bd51b10af520aafd}. It is predicted that by 2030, the number of EVs on the road will exceed 100 million, a massive increase from the 5.1 million present in 2019 \cite{bunsen2019global}. The widespread adoption of EVs will lead to a significant increase in the electricity demand on power distribution networks (PDNs) for EV charging. This poses a high risk of PDN overloading during peak-demand hours. Furthermore, the locations of EV charging stations (EVCSs) along transportation networks (TNs) can have an impact on the routes and travel times of EVs that require charging. To mitigate the negative effects of large-scale EV charging on PDNs and alleviate potential traffic congestion in TNs, proper EV routing along TNs and charging scheduling is necessary. As a result, proper coordination between the operation of EVs-infused transportation and power distribution networks is vital to ensure the seamless adoption of large-scale EVs.

In recent times, several studies have been conducted to improve the efficiency and reliability of TNs and PDNs by coordinating their operations. One such proposal put forth by the authors in \cite{7572952} suggests a centralized optimal traffic-power flow for routing of EVs in electrified TN, which has been further extended to accommodate time-varying electricity prices and traffic demands in \cite{8850088}. The traffic flow is modeled using a semi-dynamic traffic assignment technique that accounts for flow propagation between adjacent periods. Meanwhile, \cite{9626649} introduces a bilevel optimization model that allows EV aggregators to participate in day-ahead dispatch while adhering to various system operation constraints. Additionally, \cite{9951385} proposes a centralized model for a dynamic pricing strategy for integrated electricity charging and hydrogen refueling stations (IEHSs) that guides the charging and refueling decisions of different EV types and ensures the smooth operation of IEHSs, the power distribution network, and the gas network. Likewise, \cite{9816010} proposes a dynamic pricing strategy for an EV charging station that maximizes net charging profits by incorporating the behavior of EV drivers based on both an admission control scheme and a queuing model.


However, the above-mentioned studies have one main issue: a centralized operational framework. Since transportation and power distribution systems are operated by different entities, and there is no entity that has access to both TN and PDN information, implementing a centralized model of operation may not be practical due to privacy concerns and communication overhead. The TN coordinator (TNC) has the responsibility of solving the optimal traffic flow (OTF) for EVs \cite{7572952}, which involves identifying the most efficient routing and charging schedule for electric vehicles (EVs). On the other hand, the task of the PDN operator (P-DSO) is to operate the power distribution network by solving the distribution-level optimal power flow (d-OPF) problem \cite{baran1989optimal}. Since electric vehicle charging stations (EVCSs) are powered by power distribution networks (PDNs), TNs and PDNs are physically connected. The variables that are common to both OTF and d-OPF are called boundary variables (i.e., charging loads of EVCSs). If OTF and d-OPF are solved independently without coordination, boundary variables may not match, resulting in an insecure or sub-optimal performance of both systems.

After realizing this issue, researchers have proposed decentralized frameworks to coordinate TNs and PDNs with EVs. These frameworks involve obtaining the routes of EVs using the best response decomposition (BRD) algorithm \cite{7967870}, coordination of hydrogen-integrated TN and PDNs using the alternating direction method of multipliers (ADMM) \cite{9528033}, a novel optimal traffic power flow problem via an extended ADMM to analyze the spatial and temporal congestion propagation on coupled transportation power systems under congested roads, transmission lines and charging stations \cite{9802712}, multistage distributionally robust optimization model to address wind power uncertainty based on improved ADMM \cite{9718154}, coordination of the charging schedule of EVs and stochastic security-constrained unit commitment using the bender decomposition method (BDM) \cite{8429264}, augmented lagrangian alternating direction inexact newton (ALADIN) based coordination for time-varying traffic demands and inter-temporal EV charging behavior \cite{9079677}, an improved fixed-point algorithm (FPA) based on extrapolation for the spatial and temporal evolution of traffic flows \cite{9626143}. Similarly, other frameworks include stochastic user equilibrium traffic assignment using optimality condition decomposition (OCD) method \cite{9891824}, strategic pricing method to maximize the profit of EVCS owners using Karush-Kuhn-Tucker (KKT) condition \cite{9937161}, the generalized user equilibrium method for the coupled power-transportation network operation using a master and a series of subproblems (MSS) approach \cite{9891825},  a new collaborative pricing model for the power-transportation coupled network based on a variational inequality approach using an improved prediction-correction algorithm (IPCA) \cite{9748107},  the iterative column generation (CG) algorithm to explicitly describe PEVs' driving range constraints on TNs \cite{8962325}. 

To simulate the routing and charging schedule of EVs in TN, binary variables are employed, making OTF a mixed-integer program \cite{7967870}. Likewise, binary variables are used to capture the behavior of switched capacitors and voltage regulators in d-OPF, thereby rendering the d-OPF a mixed-integer program too \cite{8457301}. Nevertheless, the decentralized algorithms mentioned above exhibit at least one of the following primary issues: First, they do not provide a guarantee of optimality and convergence for mixed-integer programs (MIPs), such as OTF and d-OPF. For instance,  the BRD, ADMM, BDM, OCD, KKT, CG, and IPCA algorithms can only ensure optimality and convergence for convex problems. Second, some of these algorithms, e.g., BRD and MSS, require TNC and P-DSO to exchange a significant amount of information, which results in high investment in building communication channels and causes long communication delays.  Additionally, TNC and P-DSO may not want to share their confidential information. Third, some of these algorithms such as ALADIN and MSS are not fully decentralized as they require certain steps (such as hessian update and master subproblem) to be performed centrally (which may not be practical in a purely decentralized environment). Therefore, there is a need for a fully decentralized algorithm that facilitates the coordination of MIP problems such as OTF and d-OPF with limited information exchange only.

Recently, a promising algorithm called the SDM-GS-ALM algorithm \cite{boland2019parallelizable}, which is a combination of the simplicial decomposition method, gauss-seidel, and augmented lagrangian method, has been shown to guarantee optimality and convergence to MIP subproblems. However, when it was applied to OTF and d-OPF in our preliminary study, it proved to be computationally inefficient. Therefore, we propose a new fast and fully decentralized algorithm for the coordination of OTF and d-OPF with limited information exchange only. In summary, the proposed algorithm has the following significant contributions and benefits: 
\begin{enumerate}
    \item First, unlike ADMM \cite{boyd2011distributed}, the proposed algorithm is guaranteed to converge for MIP subproblems such as OTF and d-OPF. 
    \item Second, it requires only limited information exchange between TNC and P-DSO. The only information shared between TNC and P-DSO is boundary variables.
    \item Third, it does not require any computation by the central operator (third entity). It is worth noting that the ALADIN \cite{9079677} requires the hessian update to be performed centrally (by the central operator). In contrast, our proposed framework allows P-DSO and TNC to compute everything, including lagrangian multipliers, in a fully decentralized manner.
    \item Fourth, it is faster than the SDM-GS-ALM algorithm \cite{boland2019parallelizable}.
\end{enumerate}    
\vspace{-3mm}
\section{Problem Formulation}
\subsection{Power Distribution Network Model}
Distributed energy resources (DERs) like solar photovoltaics (PVs), and grid power, which supply EVCSs in the transportation network and other loads in the power distribution network, are considered in this paper. The resulting distribution system model is given as follows \cite{li2018micro}:
\begin{subequations}
\begin{align} 
&\left ( p_{ik} \right )^{2}+\left ( q_{ik} \right )^{2}= v_{i}\ell _{ik}\label{power_flow}\\
&v_{i}-v_{k}-2\left ( r_{ik}p_{ik}+x_{ik}q_{ik} \right ) &\nonumber\\
&+\left ( \left ( r_{ik} \right )^{2}+\left ( x_{ik} \right )^{2}  \right )\ell _{ik}=0 \label{voltage_drop}\\
&0\leq \ell _{ik}\leq \overline{\ell} _{ik} \label{thermal_limits} \\
&\left ( p_{ik} \right )^{2}+\left ( q_{ik} \right )^{2}\leq \left ( \overline{S}_{ik} \right )^{2} \label{power_flow_limits} \\
&\left ( \underline{v}_{i} \right )^{2}\leq v_{i}\leq \left ( \overline{v}_{i} \right )^{2} \label{voltage_limits} \\
&p_{i}^{\mathrm{G}}+p_{i}^{\mathrm{PV}}-p_{i}^{\mathrm{Load}}-p_{i}^{\mathrm{D}}\nonumber\\
&= \sum_{j}\left ( p_{ji}-r_{ji}\ell _{ji} \right )+\sum_{k}p_{ik} \label{power_balance1}\\
&q_{i}^{\mathrm{G}}+q_{i}^{\mathrm{PV}}-q_{i}^{\mathrm{Load}}\nonumber-q_{i}^{\mathrm{D}}\\
&= \sum_{j}\left ( q_{ji}-x_{ji}\ell _{ji} \right )+\sum_{k}q_{ik}  \label{power_balance2}\\
&q_i^{\mathrm{D}}=p_i^{\mathrm{D}}tan(\theta_i^{\mathrm{D}}). \label{EVCS_reactive_power}
\end{align} \label{Power_Constraints}
\end{subequations}

The balanced power flows are modeled using the \textit{DistFlow} model \cite{baran1989optimal}, which is expressed in equation \eqref{power_flow}. The distribution lines connecting node $i(j)$ and $k(i)$ are denoted by the index $ik(ji)$. The voltage drop on a distribution line is constrained by equation \eqref{voltage_drop}, while the thermal and power carrying limits of distribution lines are given by constraints \eqref{thermal_limits} and \eqref{power_flow_limits}, respectively. The voltage limits are specified by constraint \eqref{voltage_limits}. The nodal active and reactive power balance equations are given by constraints \eqref{power_balance1} and \eqref{power_balance2}, respectively. The constraint \eqref{EVCS_reactive_power} represents the reactive power demand due to the EVCSs, where $\theta_i^{\mathrm{D}}$ is the power factor angle of the EVCS. The description of the rest of the undefined symbols is provided in Table \ref{nomen_PDS}. Throughout the paper, \textbf{bold} symbols denote matrices/vectors of corresponding variables.

\begin{table}[htbp]
\centering
\caption{Nomenclature: PDN model}
\label{nomen_PDS}
\begin{tabular}{lll}
\hline \hline
 & Symbol & Description \\ \hline
\multirow{7}{*}{\rot{Parameters}} 
& $r_{ik},x_{ik}$ & Resistance and reactance of line $ik$. \\ \cline{2-3} 
 & $\overline{\ell} _{ik}$ & Squared of current carrying capacity of line $ik$. \\ \cline{2-3} 
 & $\overline{S}_{ik}$ & MVA limit of line $ik$. \\ \cline{2-3} 
 & $\underline{v}_{i}, \overline{v}_{i}$ & Minimum and maximum voltage limits. \\ \cline{2-3} 
 & $p_{i,t}^{\mathrm{Load}}, q_{i,t}^{\mathrm{Load}}$ & Active, reactive power demand at node $i$. \\ \cline{2-3} 
 & $p_{i}^{\mathrm{PV}}, q_{i}^{\mathrm{PV}}$ & \begin{tabular}[c]{@{}l@{}}Active, reactive power output from solar PV \\ at node $i$.\end{tabular} \\ \cline{2-3} 
 & $p_{i}^{\mathrm{D}}, q_{i}^{\mathrm{D}}$ & \begin{tabular}[c]{@{}l@{}}Active, reactive power demand of EVCS\\  at node $i$.\end{tabular} \\ \hline
\multirow{4}{*}{\rot{Variables}} & $v_{i}$ & Squared of voltage at node $i$. \\ \cline{2-3} 
 & $\ell _{ik} $ & Squared of current flow on line $ik$. \\ \cline{2-3}
 & $p_{ik},q_{ik}$ & Active, reactive power flow on line $ik$. \\ \cline{2-3} 
 & $p_{i}^{\mathrm{G}}, q_{i}^{\mathrm{G}}$ & Active and reactive power from grid at node $i$. \\ \hline \hline
\end{tabular}%
\end{table}    

\subsection{Transportation Network Model}
In this subsection, we briefly provide a description of the adopted model of the traffic flow for EVs. For a detailed understanding of the model, readers are referred to \cite{9891825}. The transportation network (TN) is represented by a directed graph $G=[N, A]$, where $N$ and $A$ refer to the set of nodes and arcs, respectively. To specify the electric vehicle (EV) travel and charging options, a path is formed by combining nodes and arcs. The nodes correspond to the starting and ending points of an arc, junctions, and charging stations for EVs, while arcs connect two nodes. Even though the traffic network will have both electric and non-electric vehicles, we consider routing and charging of only electric vehicles in this paper as in \cite{9891825}. Therefore, the traffic parameters used in this paper are non-electric vehicles discounted. The resulting model is given as follows:
\vspace{-3mm}
\begin{subequations} \label{Transportation_constraints}
\begin{align}
&\textbf{K} \delta_a^q=\textbf{I}^{rs} \label{OriDest}\\
&-M\left ( 1-\delta _a^q \right )\le e_m^q-e_n^q+d_a\beta^{rs}-E_m^q \nonumber\\
&\le M\left ( 1-\delta _a^q \right ), \forall(n,m) =a\in {A} \label{EV_energy_dyn}\\
&e_n^q-d_a\beta^{rs}\ge -M\left ( 1-\delta _a^q \right )+R^{rs}E^{rs}_{max},\nonumber\\
& \forall(n,m) =a\in {A} \label{mile_anxiety}\\
&E_m^q\le b_{m,max}, \forall m \in N \label{EV_charing}\\
&E_m^q \le \gamma _m^q E^{rs}_{max} \le e_m^q , \forall m \in N \label{EV_charging_full}\\
&e^q_{init}=E_0^{rs} \label{EV_charging_initial}\\
&E_{min}^{rs}\le e_m^q \le E_{max}^{rs}, \forall m \in N \label{EV_battery_limits}\\
&t_a=t_a^0\left ( 1+0.15\left ( {x_a}/{c_a} \right )^4 \right ), \forall a\in {A} \label{road_congestion}\\
&t_m=t_m^0(1+\alpha_mx_m), \forall m \in N \label{charge_wait_time}\\
&0\le f_q^{rs} \perp \left( C_q^{rs}-C_{rs} \right)\ge 0, \forall q \in O_{rs}\label{user_equilibrium}\\
&\sum _{q \in O_{rs}}f_q^{rs}=F^{rs}, \forall {rs} \label{traffic_flow_conservation} \\
&x_a=\sum _{rs}\sum _{q \in O_{rs}}f_q^{rs}\delta _a^q, \forall a\in {A} \label{traffic_flow_arcs}\\
&x_m=\sum _{rs}\sum _{q \in O_{rs}}f_q^{rs}\gamma _m^q, \forall m\in {N}\label{traffic_flow_nodes}\\
&C_q^{rs}=w\left( \sum_{a}^{} t_a\delta _a^q+\sum_{m}^{} \left(E_m^q/P_m^{pile} + t_m\gamma _m^q \right)\right) \nonumber\\
&+\sum_{m}^{} E_m^q \lambda_m, \forall q, \forall {rs} \label{path_cost}
\end{align}
\end{subequations}
Constraint \eqref{OriDest} specifies the starting and ending nodes of path $q$ for an origin-destination (O-D) pair $rs$. Note that the vector $\textbf{I}^{rs}$ consists of two non-zero elements of 1 and -1 at the origin and the destination node of O-D pair $rs$, respectively. Constraint \eqref{EV_energy_dyn} tracks the dynamics of the energy stored in the EVs, denoted by $e^q_m$, and the charging energy at nodes, denoted by $E^q_m$. The EV energy consumption ratio is denoted by $\beta^{rs}$, and the binary variable $\gamma_m^q$ indicates whether an EV along path $q$ is charged at node $m$. A very large positive constant $M$ is a Big-M parameter. Constraint \eqref{mile_anxiety} accounts for the range anxiety of EVs for O-D pair $rs$, represented by $R^{rs}$, where $E ^{rs} _{max}$ is the maximum energy that an EV for O-D pair $rs$ can store. Constraint \eqref{EV_charing} specifies the charging behavior of EVs, where $b_{m, max}$ is set to 0 if there is no EV charging station (EVCS) at node $m$. Constraint \eqref{EV_charging_full} ensures that the EVs are fully charged after each charging cycle, which is a realistic assumption as EV drivers tend to minimize the frequency of charging. Constraints \eqref{EV_charging_initial} and \eqref{EV_battery_limits} impose limits on the initial energy stored ($E_0^{rs}$) in the EVs and the maximum  energy (and minimum energy, $E_{min}^{rs}$) that can be stored in their batteries, respectively.

The function $t_a(x_a)$ in equation \eqref{road_congestion}, based on the commonly used Bureau of Public Roads (BPR) function \cite{9330803}, represents travel time on arc $a$, where $x_a$ is arc traffic flow. The function $t_m(x_m)$ in equation \eqref{charge_wait_time} describes the time required for charging and waiting for service at EVCSs, where $x_m$ is node traffic flow. The charging time is dependent on the rated power of the charging station, denoted by $P_m^{pile}$, while the waiting time is represented by $t^0_m$, which is fixed and is affected by the level of congestion in EV charging stations (EVCSs), estimated by $\alpha_m$. To account for congestion in the power grid at different locations, the electricity price $\lambda_m$ at an EVCS is measured using the distribution locational marginal price (DLMP) of the corresponding location. The traffic equilibrium condition is described by constraint \eqref{user_equilibrium}, which states that the travel cost $C^{rs}$ for origin-destination (O-D) pair $rs$ should be equal for all used paths and no greater than that for any unused path. This condition is also known as the Wardrop User Equilibrium (UE) Principle \cite{doi:10.1680/ipeds.1952.11259}. Constraint \eqref{traffic_flow_conservation} represents the conservation of traffic flow. In constraints \eqref{traffic_flow_arcs} and \eqref{traffic_flow_nodes}, the mathematical expression for arc flow $x_a$ and node flow $x_m$ is provided. Constraint \eqref{path_cost} provides the expression for the social cost ($C_q^{rs}$) of path $q$ for an O-D pair $rs$, which includes the cost of travel time, charging time for EVs, and the charging cost for EVs. The description of the rest of the undefined symbols is provided in Table \ref{nomen_OTF}.

Note that this paper considers the static Traffic Assignment Problem (TAP), which is the foundation for dynamic traffic assignment. However, the dynamic TAP is more complex than the static TAP, and there is no universally accepted model available for it \cite{9891825}. Moreover, the existing literature that considers the coordination of PDN and TN with EVs mostly uses static TAP for optimal coordination \cite{9891825}. Nonetheless, the proposed framework and algorithms would work for a dynamic TAP if it was available. 

\begin{table}[htbp]
\centering
\caption{Nomenclature: TN model}
\label{nomen_OTF}
\begin{tabular}{lll}
\hline \hline
 & Symbol & Description \\ \hline
\multirow{2}{*}{\rot{Indices}} & \begin{tabular}[c]{@{}l@{}}$m/n$,$a$, \\ $rs$, $q$\end{tabular} & \begin{tabular}[c]{@{}l@{}}Index for nodes, arcs, O-D pairs, and path in \\ the traffic network\end{tabular} \\ \cline{2-3} 
 & $evcs$ & Superscript for EVCS \\ \hline
\multirow{7}{*}{\rot{Parameters}} 
 & $\textbf{K}$ & Node-arc incidence matrix \\ \cline{2-3} 
 & $\delta_a^q$ & The path-arc incidence matrix \\ \cline{2-3} 
 & $d _{a}^{}$ & Travelling distance of arc $a$ \\ \cline{2-3} 
 & $t_a^0,c_a$ & Parameters for travel time function of arc $a$ \\ \cline{2-3} 
 & $P_{m,max}^{evcs}$ & Maximum power output of EVCS at node $m$ \\ \cline{2-3} 
 & $w$ & Weighting factor from time to monetary cost \\ \cline{2-3} 
 & $F ^{rs}$ & Travel demand of O-D pair $rs$ \\ \hline \hline
\end{tabular}%
\end{table}
\vspace{-3mm}
\subsection{Coupling between PDN and TN}
From a system-level perspective, on-road fast charging stations would simultaneously impact vehicle routing in the transportation system and load flows in the distribution system, therefore, tightly coupling the two systems. Mathematically, the coupling between PDN and TN can be captured as follows:
\begin{subequations}
\begin{align}
&p_m^{\mathrm{T}}=\sum _{rs}\sum _{q \in O_{rs}}f_q^{rs}E_m^q, \forall m\in {N} \label{coupling_constraint}\\
&0\le p_m^{\mathrm{T}} \le p_{m,max}^{\mathrm{T}}, \forall m\in {N} \label{EVCS_power_limit}
\end{align}  \label{coupling_PQ}
\end{subequations}
where $p_m^{\mathrm{T}}$ is the power consumption of EVs in TN and acts as a power demand to PDN. The constraint \eqref{coupling_constraint} is the active power consumed at the EVCSs while \eqref{EVCS_power_limit} provides the capacity limit of EVCSs.
\vspace{-4mm}
\subsection{Convexification and linearization of non-linear constraints}
The PDN model \eqref{Power_Constraints} is in non-linear programming (NLP) form due to the non-linear constraint \eqref{power_flow}. Similarly, the TN model \eqref{Transportation_constraints} is in mixed-integer non-linear programming (MINLP) form due to the non-linear BPR function \eqref{road_congestion} and bilinear traffic equilibrium constraint \eqref{user_equilibrium}. In addition, the coupling constraint \eqref{coupling_constraint} is non-linear due to the multiplication of two continuous variables. In reference \cite{li2018micro}, it is demonstrated that the solution of the mixed-integer convex program (MICP) form of the d-OPF model \eqref{Power_Constraints} corresponds to the solutions of the MINLP form, but with a reduced computational burden. Therefore, in this paper, we employ convex hull relaxation \cite{li2018micro} and piecewise linearization techniques \cite{beale1970special}, as MICP models are theoretically simpler to coordinate than MINLP models. It is worth noting that the convex hull relaxations and  piecewise linearizations are adopted from \cite{li2018micro} and \cite{beale1970special}, respectively. Therefore, for their accuracy, readers are referred to \cite{li2018micro} and \cite{beale1970special}. The convex hulls relaxation of \eqref{power_flow} is given as follows:
\begin{align} \label{CH_power}
\left\lbrace 
\begin{array}{l} 
p_{ik}^2+q_{ik}^2 \leq v_{i}\ell _{ik} \\ \overline{S}_{ik}^2 {v}_i+\underline{ {v}}_i\overline{ {v}}_i\ {\ell}_{ik} \leq \overline{S}_{ik}^2(\underline{ {v}}_i+\overline{ {v}}_i). 
\end{array}\right. 
\end{align}
The BPR function constraint \eqref{road_congestion} is a non-linear constraint, which is modeled by a piecewise linearization technique using a special ordered set of type-2 (SOS2) variables \cite{beale1970special}. In piecewise linearization, the non-linear function is divided into a number of segments, and each segment is represented by a linear function. The piecewise linearization of the BPR function constraint \eqref{road_congestion} is given as follows:
\begin{subequations} \label{road_congestion_LN}
\begin{align} 
&x_a=\sum _n x_a^ny_a^n\\
&t_a=\sum _n t_a^n\left ( x_a^n \right )y_a^n\\
&\sum  _n y_a^n=1\\
&0 \le \left \{y_a^n, \forall n  \right \} \in \text{SOS2},
\end{align}
\end{subequations}
where $x_a^n$ and $t_a^n$ denote the values of $x_a$ and $t_a$, respectively, while $y_a^n$ is a vector of SOS2 variables, in which at most two adjacent variables can be nonzero. For a detailed understanding of the linearization technique, please refer to \cite{beale1970special}.

The user equilibrium constraint \eqref{user_equilibrium} is a bilinear constraint, which is linearized using the Big-M method as follows:
\begin{subequations} \label{user_equilibrium_LN}
\begin{align}
&0 \le f_q^{rs} \le M \nu _q^{rs}, \forall q , rs\\
&0 \le C_q^{rs}-C^{rs} \le M\left ( 1-\nu _q^{rs} \right ), \forall q , \forall rs,
\end{align}
\end{subequations}
where, $\nu _q^{rs}$ is a binary variable for each path.

The coupling constraint \eqref{coupling_constraint} is also a non-linear constraint. The same piecewise linearization technique, used in \eqref{road_congestion_LN}, is applicable for the linearization of \eqref{coupling_constraint}. Owing to the page limit, the detailed modeling is omitted.

Finally, the convex constraints set of PDN, denoted by $\mathcal{X}_{\mathrm{p}}$, is defined as follows:
\begin{equation} \label{PDN_set}
    \mathcal{X}_{\mathrm{p}}:=\left \{ \text{\eqref{voltage_drop}-\eqref{power_balance2}, \eqref{CH_power}} \right \}
\end{equation}
while the mixed-integer convex constraints set of TN, denoted by $\mathcal{X}_{\mathrm{v}}$, is defined as follows:
\begin{equation} \label{TN_set}
    \mathcal{X}_{\mathrm{v}}:=\left \{ \text{\eqref{OriDest}-\eqref{EV_battery_limits}, \eqref{charge_wait_time}, \eqref{traffic_flow_conservation}-\eqref{path_cost}, 
    \eqref{coupling_PQ}, 
    \eqref{road_congestion_LN}, \eqref{user_equilibrium_LN}} \right \}.
\end{equation} 

\subsection{Coordinated d-OPF-OTF formulation under the decentralized framework}
This subsection presents the formulations of the coordinated d-OPF-OTF problem under the decentralized framework. In our proposed framework, the power operator solves the following d-OPF subproblem:
\begin{subequations}\label{CDEWN0P}
\begin{align}
    \text{(PO.) }\min_{\bm{x_{\mathrm{p}}}, \bm{\alpha_{\mathrm{p}}}} &~f_{\mathrm{p}}\left( \bm{x_{\mathrm{p}}} \right):= \sum _i c_i^{\mathrm{G}}p_i^{\mathrm{G}}=\bm{c^{\mathrm{G}}}\bm{p^{\mathrm{G}}}\label{power_obj}\\
    \text {s.t. }
    &~\bm{p^{\mathrm{D}}}=\boldsymbol{z}\label{consensusC1P}\\
    &~\bm{x_{\mathrm{p}}} \in \mathcal{X}_{\mathrm{p}}, \bm{\alpha_{\mathrm{p}}} \in \{0,1\},
\end{align}
\end{subequations}
where $c^{\mathrm{G}}$ is electricity price (can be interpreted as the transmission system's (grid) locational marginal prices (LMPs)) and ${p^{\mathrm{G}}}$ is electric power purchased by PDN from the grid. As such, the objective function \eqref{power_obj} minimizes the power purchased from the grid. $\bm{p^{\mathrm{D}}}$ represents the power demand due to EVCSs in the d-OPF model. Auxiliary variable $\bm{z}$ is utilized to facilitate the decentralized operation. All the decision variables of the PDN constraint set \eqref{PDN_set} are collectively referred to as $\bm{x_{\mathrm{p}}}$ while $\bm{\alpha_{\mathrm{p}}}$ collectively represents the integer decision variables. Even though there are no integer (including binary) variables present in the PDN constraint set \eqref{PDN_set}, we have introduced binary variables $\bm{\alpha_{\mathrm{p}}}$ in the compact form for the generalization (for the potential future adoption) of the proposed algorithm.

Similarly, the transportation coordinator solves the following OTF subproblem:
\begin{subequations}\label{CDEWN0W}
\begin{align}
   \text{(TO.) }\min_{\bm{x_{\mathrm{v}}},\bm{\alpha_{\mathrm{v}}}} &~f_{\mathrm{v}}\left( \bm{x_{\mathrm{v}}} \right):=\sum _{rs} \sum _q f_q^{rs}C_q^{rs} \nonumber\\
   &=\sum _{rs} F^{rs}C^{rs}=\boldsymbol{F}\boldsymbol{C} \label{transportation_obj}\\
    \text {s.t. }
    &~\bm{p^{\mathrm{\mathrm{T}}}}=\boldsymbol{z} \label{consensusC1W}\\
&~\bm{x_{\mathrm{v}}}\in {\mathcal{X}_{\mathrm{v}}}, \bm{\alpha_{\mathrm{v}}} \in \{0,1\},
\end{align}
\end{subequations}
where $\bm{x_{\mathrm{v}}}$ collectively represents all the decision variables while $\bm{\alpha_{\mathrm{v}}}$ collectively represents the integer decision variables of the TN constraint set \eqref{TN_set}. And, $\bm{p^{\mathrm{\mathrm{T}}}}$ represents the power consumed by EVCSs in OTF. The objective function of the transportation subproblem \eqref{transportation_obj} minimizes the social cost of the transportation sector (i.e., time and energy consumption cost) \cite{9891825}.

The decentralized formulation, i.e., \eqref{CDEWN0P} and \eqref{CDEWN0W} has one significant advantage: it does not require any entity with access to both ${\mathcal{X}_{\mathrm{p}}}$ and ${\mathcal{X}_{\mathrm{v}}}$. It is important to note that there does not exist any entity that has access to both PDN and TN information. Therefore, the proposed decentralized formulation of coordinated d-OPF-OTF provides a real-world-compatible framework for the coordination of PDN and TN. Nonetheless, it can be observed from \eqref{consensusC1P} and \eqref{consensusC1W} that the two subproblems are still coupled through $\boldsymbol{z}$ as EVCS powers $\bm{\bm{p^{\mathrm{T}}}}$ in a TN act as a power demand $\bm{p^{\mathrm{D}}}$ in a PDN. If two models are solved independently without being coordinated by a proper decentralized algorithm, the boundary variables, i.e., $\bm{\bm{p^{\mathrm{D}}}}$ and $\bm{\bm{p^{\mathrm{T}}}}$ may not match with each other, which will result in increased cost or insecure operation of PDN. Therefore, in the next section, we introduce a novel decentralized algorithm that allows TNC and P-DSO to solve two subproblems separately but coordinately, with the guarantee of boundary variables matching. 
\vspace{-3mm}
\section{Decentralized Algorithm}
As mentioned in the introduction section, the most existing decentralized or distributed optimization algorithms used to coordinate d-OPF \eqref{CDEWN0P} and OTF \eqref{CDEWN0W} are not guaranteed to converge and be optimal for MIP problems or are not fully decentralized, or require a significant amount of information sharing. Moreover, the SDM-GS-ALM algorithm \cite{boland2019parallelizable} was proved to be computationally inefficient in our preliminary study. Therefore, this paper proposes a new fast decentralized optimization algorithm that is capable of handling mixed-integer subproblems such as d-OPF \eqref{CDEWN0P} and OTF \eqref{CDEWN0W} in a fully decentralized manner. The faster convergence comes from adaptively selecting inner loop iterations in the computation of lagrangian upper bounds instead of pre-fixed inner loop iteration as in SDM-GS-ALM algorithm \cite{boland2019parallelizable}. This avoids setting 1) high inner loop iterations, potentially saving time and iterations 2) low inner loop iterations, potentially hampering convergence properties of the algorithm. We first provide an overview of the proposed variable inner loop selection (VILS) based decentralized algorithm in Subsection III-A and then prove the optimality and convergence in the second subsection. 


\subsection{Algorithm Overview}

\begin{algorithm}[H]
\caption{VILS Decentralized Algorithm}
\label{alg:euclid2}
\begin{algorithmic}[1]
\\\textbf{Parameters initialization: }
\begin{enumerate}
    \item \textbf{Parameters selection: }Choose the outer loop convergence tolerance $\epsilon$, lagrangian upper bound tolerance (inner loop) $\epsilon_{\mathrm{u}}$, penalty parameter $\gamma$, outer loop iteration limit $K$. \\
    \textbf{Starting point: }Starting points for auxiliary variable ${\bm{z}}$, binary variables $\bm{\alpha_{\mathrm{p}}}$ and $\bm{\alpha_{\mathrm{v}}}$, lagrangian multipliers $\boldsymbol{\lambda _{\mathrm{p}}}$ and $\boldsymbol{\lambda _{\mathrm{v}}}$, and lagrangian lower bounds $\check {\varphi}_{\mathrm{p}}$ and $\check {\varphi}_{\mathrm{v}}$ are assigned.
\end{enumerate}

\\\textbf{Iteration initialization: } Set $LR^k:=\{0\}$ and $\Delta LR^k:=\{\text{a large number}\}$, and $ \{\boldsymbol z, \boldsymbol{\lambda _{\mathrm{p}}},\boldsymbol{\lambda _{\mathrm{v}}}, \bm{\alpha_{\mathrm{p}}}, \bm{\alpha_{\mathrm{v}}}, \check {\varphi}_{\mathrm{p}}, \check {\varphi}_{\mathrm{v}}\}^k := \{\bm{z},\boldsymbol{\lambda _{\mathrm{p}}},\boldsymbol{\lambda _{\mathrm{v}}}, \bm{\alpha_{\mathrm{p}}}, \bm{\alpha_{\mathrm{v}}},\check {\varphi}_{\mathrm{p}}, \check {\varphi}_{\mathrm{v}}\}^{k-1}$.

\\\textbf{Lagrangian upper bound computation: }
While $\Delta LR^k \ge \epsilon_u$, repeat the following \eqref{repeatthis}:
\begin{subequations} \label{repeatthis}
\begin{align}
&LR^k_{\mathrm{p}}, \bm{x_{\mathrm{p}}}^k , \bm{p}^{\bm{\mathrm{D}}(k)}\gets \min _{\bm{x_{\mathrm{p}}}, \bm{p^{\mathrm{D}}}} \{ L^{\mathrm{p}}_\gamma\left( \bm{x_{\mathrm{p}}} ,\bm{p}^{\bm{\mathrm{D}}},\bm{z}^k, \bm{\lambda_{\mathrm{p}}}^k \right) : \nonumber \\
&  \bm{\alpha_{\mathrm{p}}} \in \bm{\alpha_{\mathrm{p}}}^k, \bm{x_{\mathrm{p}}} \in {\mathcal{X}_{\mathrm{p}}}  \} \label{LR_upper_power}\\
&LR^k_{\mathrm{v}}, \bm{x_{\mathrm{v}}}^k , \bm{p}^{\bm{\mathrm{T}}(k)} \gets \min _{\bm{x_{\mathrm{v}}}, \bm{p^{\mathrm{T}}}} \{ L^{\mathrm{v}}_\gamma\left( \bm{x_{\mathrm{v}}},\bm{p}^{\bm{\mathrm{T}}},\bm{z}^k, \bm{\lambda_{\mathrm{v}}}^k \right) : \nonumber \\
& \bm{\alpha_{\mathrm{v}}} \in \bm{\alpha_{\mathrm{v}}}^k, \bm{x_{\mathrm{v}}} \in {\mathcal{X}_{\mathrm{v}}} \}\label{LR_upper_transportation}\\
&\bm{z}^k \gets \min _{\bm{z}} \left\{ \left\| \bm{z}-\bm{p}^{\bm{\mathrm{D}}(k)} \right\|_2^2+\left\| \bm{p}^{\bm{\mathrm{T}}(k)}-\bm{z} \right\|_2^2 \right\} \label{zupdate}\\
&LR^k \gets LR^k \cup  \{(LR^k_p+LR^k_v)\} \label{LR_function_value_ set}\\
&\Delta LR^k \gets LR^k_{end}-LR^k_{end-1} \label{LR_function_difference} 
\end{align}
\end{subequations}
And, obtain the lagrangian upper bounds as follows:
\begin{subequations} \label{Upper_bounds}
\begin{align}
&\hat{\varphi}_{\mathrm{p}}^{k} \gets LR^k_{\mathrm{p}}+\frac{\gamma}{2}\left\| \bm{z}^k-\bm{p}^{\bm{\mathrm{D}}(k)} \right\|_2^2\\
&\hat{\varphi}_{\mathrm{v}}^{k} \gets LR^k_{\mathrm{v}}+\frac{\gamma}{2}\left\| \bm{p}^{\bm{\mathrm{T}}(k)}-\bm{z}^k \right\|_2^2
    \end{align}
\end{subequations}

\\\textbf{Convergence check: }If $(\hat{\varphi}_{\mathrm{p}}^{k}+\hat{\varphi}_{\mathrm{v}}^{k})-(\check{\varphi}_{\mathrm{p}}^{k}+\check{\varphi}_{\mathrm{v}}^{k})\le \epsilon$, Stop: $\left( \bm{x_{\mathrm{p}}}^k, \bm{x_{\mathrm{v}}}^k,\boldsymbol{z}^k,\boldsymbol{\lambda _{\mathrm{p}}}^k,\boldsymbol{\lambda _{\mathrm{v}}}^k,\check{\varphi}_{\mathrm{p}}^k, \check{\varphi}_{\mathrm{v}}^k \right)$ is the solution. Otherwise, continue.

\\\textbf{Lagrangian lower bound computation: }
Compute the intermediate lagrangian lower bound as follows:
\begin{subequations}
\begin{align}
&\tilde{\varphi}_{\mathrm{p}}, \hat{ \bm{x}}_{\bm{\mathrm{p}}}, \bm{\alpha _{\mathrm{p}}}^k \gets \check{\varphi}_{\mathrm{p}} \left( \boldsymbol{\lambda _{\mathrm{p}}}^k+\gamma\left( \boldsymbol{z}^k-\boldsymbol{p}^{\boldsymbol{\mathrm{D}}(k)} \right) \right) \label{tilde1} \\
&\tilde{\varphi}_{\mathrm{v}}, \hat{ \bm{x}}_{\bm{\mathrm{v}}},\bm{ \alpha _{\mathrm{v}}}^k \gets \check{\varphi}_{\mathrm{v}} \left( \boldsymbol{\lambda _{\mathrm{v}}}^k+\gamma\left( \boldsymbol{p}^{\boldsymbol{\mathrm{T}}(k)}-\boldsymbol{z}^k \right) \right) \label{tilde2}
\end{align}
\end{subequations}

\algstore{myalg}
\end{algorithmic}
\end{algorithm}
\begin{algorithm}                     
\begin{algorithmic} [1]               
\algrestore{myalg}

\\\textbf{Lagranian lower bound quality check: } The intermediate lagrangian lower bound passes the quality check and iteration is declared \textit{forward} iteration if the following inequality holds:
\begin{equation}
(\hat{\varphi}_{\mathrm{p}}^{k}+\hat{\varphi}_{\mathrm{v}}^{k}) \ge (\tilde{\varphi}_{\mathrm{p}}^{k}+\tilde{\varphi}_{\mathrm{v}}^{k}) \ge (\check{\varphi}_{\mathrm{p}}^{k}+\check{\varphi}_{\mathrm{v}}^{k}). \label{declare_iterationll}
\end{equation}
Perform the dual variables updates and keep the lagrangian lower bounds if the iteration is declared \textit{forward}:
\begin{subequations}
\begin{align}
&\boldsymbol{\lambda _{\mathrm{p}}}^k \gets \boldsymbol{\lambda _{\mathrm{p}}}^k+\gamma\left( \boldsymbol{z}^k-\boldsymbol{p}^{\boldsymbol{\mathrm{D}}(k)} \right)\label{lambda1update}\\
&\boldsymbol{\lambda _{\mathrm{v}}}^k \gets \boldsymbol{\lambda _{\mathrm{v}}}^k+\gamma\left( \boldsymbol{p}^{\boldsymbol{\mathrm{T}}(k)}-\boldsymbol{z}^k \right)\label{lambda2update} \\
&\check{\varphi}_{\mathrm{p}}^{k} \gets \tilde{\varphi}_{\mathrm{p}}^{k}\label{lower1update}\\  &\check{\varphi}_{\mathrm{v}}^{k} \gets \tilde{\varphi}_{\mathrm{v}}^{k}\label{lower2update}
\end{align}
\end{subequations}
Otherwise, the iteration is declared \textit{neutral}: Algorithm continues without updates.

\\\textbf{Loop: }Set $k:=k+1$ and go back to Step 2.
\end{algorithmic}
\end{algorithm}

The key steps of the proposed VILS algorithm are provided in Algorithm 1. The algorithm is initialized by assigning parameters in Step 1. Moreover, the starting points (i.e., iteration $0$) for the auxiliary variable ${\bm{z}}$, binary variables $\bm{\alpha_{\mathrm{p}}}$ and $\bm{\alpha_{\mathrm{v}}}$, lagrangian multipliers $\boldsymbol{\lambda _{\mathrm{p}}}^k$ and $\boldsymbol{\lambda _{\mathrm{v}}}^k$, and lagrangian lower bounds $\check {\varphi}_{\mathrm{p}}$ and $\check {\varphi}_{\mathrm{v}}$ are assigned. Note that $\bm{\alpha_{\mathrm{p}}}$ and $\bm{\alpha_{\mathrm{v}}}$ collectively represent the binary variables of PDN and TN subproblems, respectively.  For the initial values of auxiliary variable ${\bm{z}}$ and lagrangian multipliers $\boldsymbol{\lambda _{\mathrm{p}}}^k$ and $\boldsymbol{\lambda _{\mathrm{v}}}^k$, we can use zero. For the initial values of binary variables $\bm{\alpha_{\mathrm{p}}}$ and $\bm{\alpha_{\mathrm{v}}}$, we can use any feasible solution. For the lagrangian lower bounds $\check {\varphi}_{\mathrm{p}}$ and $\check {\varphi}_{\mathrm{v}}$, we can use any small negative number. Note that Step 2 to Step 8 constitutes the outer loop while \eqref{repeatthis} in Step 3 is the inner loop.  

For any current iteration $k$, the first element of the lagrangian function value set $LR^k$ is set to 0 while the difference of the lagrangian function value $\Delta LR^k$ is initially set to a large number in Step 2. In addition, the initial values of auxiliary variable ${\bm{z}}$, binary variables $\bm{\alpha_{\mathrm{p}}}$ and $\bm{\alpha_{\mathrm{v}}}$, lagrangian multipliers $\boldsymbol{\lambda _{\mathrm{p}}}^k$ and $\boldsymbol{\lambda _{\mathrm{v}}}^k$, and lagrangian lower bounds $\check {\varphi}_{\mathrm{p}}$ and $\check {\varphi}_{\mathrm{v}}$ are set to that of the previous iteration $k-1$.

The $L_{\gamma}^{\mathrm{p}}$ in \eqref{LR_upper_power} and $L_{\gamma}^{\mathrm{v}}$ in \eqref{LR_upper_transportation}  have the following detailed expressions in Step 3:
\begin{subequations} \label{Upper_LR_bound}
\begin{align}
&L_{\gamma}^{\mathrm{p}}=\bm{c^{\mathrm{G}}}\bm{p^{\mathrm{G}}}- (\boldsymbol{\lambda _{\mathrm{p}}})^\top{\bm{p^{\mathrm{D}}}} + \frac{\gamma }{2} \left\| {{\boldsymbol{z}}-\bm{p^{\mathrm{D}}} } \right\|_2^2, \label{LR_of_power}\\
&L_{\gamma}^{\mathrm{v}}=\boldsymbol{F}\boldsymbol{C}+(\boldsymbol{\lambda _{\mathrm{v}}})^\top{\bm{\bm{p^{\mathrm{T}}}}} + \frac{\gamma }{2} \left\| {\bm{\bm{p^{\mathrm{T}}}} - {\boldsymbol{z}}} \right\|_2^2. \label{LR_of_transportation}
\end{align}
\end{subequations} 
Note that \eqref{LR_of_power} and \eqref{LR_of_transportation}, the augmented Lagrangian relaxations of \eqref{CDEWN0P} and \eqref{CDEWN0W}, respectively, are computed in parallel by P-DSO and TNC, respectively. In \eqref{LR_of_power} and \eqref{LR_of_transportation}, binary variables are fixed so that PDN and TN sub-problems are continuous. The binary variables are fixed from the solutions of the previous iteration of MIP subproblems in Step 4. Moreover, the auxiliary variable $\bm{z}$ is computed as in \eqref{zupdate}. The auxiliary variable update \eqref{zupdate} can be assigned to either of the operators (in our study, we assign it to the TN coordinator) as the only information shared is the boundary variables from both networks. Moreover, the lagrangian function value set $LR^k$ is updated as in \eqref{LR_function_value_ set} while the difference of the lagrangian function value $\Delta LR^k$ is updated as in \eqref{LR_function_difference}, where two most recent elements of $LR^k$ are utilized (subscript $end$ represents the most recent element). Finally, the lagrangian upper bounds $\hat{\varphi}_{\mathrm{p}}$ and $\hat{\varphi}_{\mathrm{v}}$ are computed as in \eqref{Upper_bounds} in Step 3. Note that Algorithm 1 is said to converge if the difference of Lagrangian bounds $(\hat{\varphi}_{\mathrm{p}}+\hat{\varphi}_{\mathrm{v}})-(\check{\varphi}_{\mathrm{p}}+\check{\varphi}_{\mathrm{v}}))$ is within the limit of tolerance, as stated in Step 4. In this paper, the proposed VILS algorithm is used to coordinate the MICP subproblems. Therefore, it converges to the global optimal solution of the centralized implementation of MICP subproblems. 

The $\check\varphi_{\mathrm{p}}$ and $\check\varphi_{\mathrm{v}}$ (\eqref{tilde1} and \eqref{tilde2} respectively) used to obtain the intermediate lagrangian lower bounds $\tilde\varphi_{\mathrm{p}}$ and $\tilde\varphi_{\mathrm{v}}$ in Step 5 are 
 also computed in parallel by P-DSO and TNC, respectively, and given as follows:
\begin{subequations} \label{VILS_lower}
\begin{align}
     &\check\varphi_{\mathrm{p}}\left( \boldsymbol{\lambda _{\mathrm{p}}}^k \right)=\min_{\bm{x_{\mathrm{p}},\bm{p^{\mathrm{D}}}},\bm{\alpha _{\mathrm{p}}}}\left\{ \bm{c^{\mathrm{G}}}\bm{p^{\mathrm{G}}} - (\boldsymbol{\lambda _{\mathrm{p}}}^k)^\top  \bm{p^{\mathrm{D}}}: \bm{x_{\mathrm{p}}} \in {\mathcal{X}_{\mathrm{p}}}  \right\},\\
     &\check\varphi_{\mathrm{v}}\left( \boldsymbol{\lambda _{\mathrm{v}}} ^k\right)=\min_{\bm{x_{\mathrm{v}}},\bm{p^{\mathrm{T}}},\bm{\alpha _{\mathrm{v}}}}\left\{\boldsymbol{F}\boldsymbol{C}+(\boldsymbol{\lambda _{\mathrm{v}}}^k)^\top   \bm{p^{\mathrm{T}}}: \bm{x_{\mathrm{v}}}  \in  {\mathcal{X}_{\mathrm{v}}} \right\}.
\end{align}
\end{subequations}
Note that the binary variables are not fixed in Step 5, although they are fixed in Step 3. The intermediate lagrangian lower bounds ($\check\varphi_{\mathrm{p}}$ and $\check\varphi_{\mathrm{v}}$) computed in Step 5 go through a quality check in Step 6. If the intermediate lagrangian lower bounds calculated in Step 5 are greater than the previously calculated lower bounds ($\check\varphi_{\mathrm{p}}^k$ and $\check\varphi_{\mathrm{v}}^k$) and smaller than the current upper bounds ($\hat\varphi_{\mathrm{p}}^k$ and $\hat\varphi_{\mathrm{v}}^k$) as stipulated in \eqref{declare_iterationll}, the intermediate lower bounds pass the quality check (iteration is declared \textit{forward}), and lagrangian multipliers are updated in a decentralized manner as in \eqref{lambda1update} and \eqref{lambda2update}. Moreover, the lagrangian lower bounds ($\check\varphi_{\mathrm{p}}^k$ and $\check\varphi_{\mathrm{v}}^k$) are also updated as in \eqref{lower1update} and \eqref{lower2update}. Otherwise, the algorithm continues without updates.
\vspace{-3mm}
\subsection{Optimality and Convergence}
It is worth noting that the following key features of the proposed algorithm enable it to converge to the global optimal solution:
\begin{enumerate}
    \item The Lagrangian upper bound, computed using continuous subproblems as stated in Step 3, is a global upper bound.
    \item For the computation of the Lagrangian lower bound, mixed-integer convex sets $\mathcal{X}_{\mathrm{p}}$ and $\mathcal{X}_{\mathrm{v}}$ are utilized. However, as mentioned in \cite{lubin2013parallelizing}, minimizing linear objective function over mixed-integer convex sets $\mathcal{X}_{\mathrm{p}}$ and $\mathcal{X}_{\mathrm{v}}$ is equivalent to minimizing linear objective function over convex hulls sets, $\text{CH}(\mathcal{X}_{\mathrm{p}})$ and $\text{CH}(\mathcal{X}_{\mathrm{v}})$. Therefore, the lagrangian lower bound obtained is a global lower bound.
    \item The Lagrangian multipliers are computed based on the values of variables obtained from continuous subproblems; see Step 6.
\end{enumerate}

For the convergence study of the proposed VILS decentralized algorithm, we make the following two mild assumptions:
\begin{enumerate}
    \item The global optimal solution of the coordinated d-OPF-OTF is unique. This is a practical assumption, as no two devices or systems are identical.
    \item The objective function is linear. This holds true in our study.
\end{enumerate}

Finally, based on the proposed VILS  decentralized algorithm, the following theorem is established:

\noindent
{\bf Theorem: } The sequence $\left\{ (\boldsymbol{x}^k,\boldsymbol{z}^k) \right\}$ generated by the Algorithm 1 converges to the global optimal solution of MICP d-OPF and OTF subproblems as $k \rightarrow \infty$.

{\bf Proof: } We introduce the following definitions for brevity and conciseness:
\begin{subequations}
\begin{align}
&{L_{\gamma}}:={L^{\mathrm{p}}_{\gamma}}+{L^{\mathrm{v}}_{\gamma}},\nonumber\\
& \varphi^{\mathrm{c}}\left( {\boldsymbol{\lambda}}  \right):=\varphi^{\mathrm{c}}_{\mathrm{p}}\left( \boldsymbol{\lambda _{\mathrm{p}}}\right) + \varphi^{\mathrm{c}}_{\mathrm{v}}\left( \boldsymbol{\lambda _{\mathrm{v}}}\right), \nonumber\\
&\varphi^{\mathrm{c}}_{\mathrm{p}}\left( \boldsymbol{\lambda _{\mathrm{p}}}^k \right):=\min_{\bm{x_{\mathrm{p}},\bm{p^{\mathrm{D}}}}}\left\{ \bm{c^{\mathrm{G}}}\bm{p^{\mathrm{G}}} - (\boldsymbol{\lambda _{\mathrm{p}}}^k)^\top  \bm{p^{\mathrm{D}}}: \bm{x_{\mathrm{p}}} \in \text{CH}({\mathcal{X}_{\mathrm{p}}})  \right\}, \nonumber\\
&\varphi^{\mathrm{c}}_{\mathrm{v}}\left( \boldsymbol{\lambda _{\mathrm{v}}} ^k\right):=\min_{\bm{x_{\mathrm{v}}},\bm{p^{\mathrm{T}}}}\left\{\boldsymbol{F}\boldsymbol{C}+(\boldsymbol{\lambda _{\mathrm{v}}}^k)^\top   \bm{p^{\mathrm{T}}}: \bm{x_{\mathrm{v}}}  \in  \text{CH}({\mathcal{X}_{\mathrm{v}}}) \right\}, \nonumber\\
&\text{CH}(.):= \text{Convex Hulls}, \nonumber\\
& \check\varphi\left(  \boldsymbol{\lambda} \right):=\check\varphi_{\mathrm{p}}\left( \boldsymbol{\lambda _{\mathrm{p}}} \right) + \check\varphi_{\mathrm{v}}\left( \boldsymbol{\lambda _{\mathrm{v}}} \right), \nonumber\\
& \hat\varphi\left(\bm{x},\bm{z}, \boldsymbol{\lambda}\right):=\hat\varphi_{\mathrm{p}}\left(\bm{x_{\mathrm{p}}},\bm{z}, \boldsymbol{\lambda _{\mathrm{p}}} \right)+\hat\varphi_{\mathrm{v}}\left(\bm{x_{\mathrm{v}}},\bm{z}, \boldsymbol{\lambda _{\mathrm{v}}} \right) \nonumber,\\
&f(\bm{x},\bm{z}):=f_{\mathrm{p}}+f_{\mathrm{v}}=\eqref{power_obj}+\eqref{transportation_obj}, \nonumber \\
& \mathcal{X}:=\mathcal{X}_{\mathrm{p}} \cup \mathcal{X}_{\mathrm{v}},\nonumber \\
& \boldsymbol{x}^k:=(\boldsymbol{x_{\mathrm{p}}}^k, \boldsymbol{x_{\mathrm{v}}}^k)  \left(\text{vector concatenation}\right), \nonumber\\
&\left( \bm{x} - \boldsymbol{z} \right):=\left( \left( \boldsymbol{{p}^{\mathrm{T}}} - \boldsymbol{z} \right) , \left( \boldsymbol{z} - \boldsymbol{{p}^{\mathrm{D}}}  \right) \right). \nonumber
\end{align}
\end{subequations}
The convergence condition at $\bm{x} \in \mathcal{X}$ 
for a limit point $(\bar{\bm{x}},\bar{\bm{z}})$ of the sequence $\left\{ (\bm{x}^{k},\bm{z}^{k}) \right\}$ is defined as \cite{boland2019parallelizable}: 
\begin{align}
&f_{\bm{x}}'(\bm{x},\bm{z};s) \ge 0 \quad \text {for all}\; s \in \mathcal{X} - \left\{ \bm{x} \right\}, \label{convergence_condition} 
\end{align}
where $f_{\bm{x}}'(\bm{x},\bm{z};s) =lim _{\beta \to 0} \frac{f(\bm{x}+\beta s, \bm{z})-f(\bm{x},\bm{z})}{\beta}$ for some $\beta$.\\
\noindent
The Direction Related Assumption is given as follows: for any iteration $k$, $s^k$ is chosen such that $\bm{x}^{k}+s^k \in \mathcal{X}$ and $f_{\bm{x}}'(\bm{x},\bm{z};s) \ge 0$. Note that $s^k$ is a gradient of $\bm{x}^{k}$.

The proof has two parts. Part 1 proves the convergence, while Part 2 verifies the optimality.

\noindent

{\it Part 1: The sequence  $\left\{ (\bm{x}^{k},\bm{z}^{k}) \right\}$ generated by Algorithm 1 always converges to the limit point $(\bar{\bm{x}},\bar{\bm{z}})$.} \\
Here, we prove that the limit point $(\bar{\bm{x}},\bar{\bm{z}})$ of the sequence  $\left\{ (\bm{x}^{k},\bm{z}^{k}) \right\}$ of feasible solutions to the problems \eqref{CDEWN0P} and \eqref{CDEWN0W}  satisfies the convergence condition \eqref{convergence_condition}.
According to the Armijo rule \cite{quarteroni2010numerical}, we have 
\begin{align} 
\frac{f\left( \bm{x}^{k} + \beta ^k s^k,\bm{z}^{k}\right) - f\left( \bm{x}^{k},\bm{z}^{k}\right) }{\beta ^k} \le \sigma f_{\bm{x}}'\left( \bm{x}^{k},\bm{z}^{k} ; s^k\right) 
\end{align} for any $\sigma \in (0,1)$.
Note that $\beta ^k$ is the step length of the Armijo rule \cite{quarteroni2010numerical}. As $f_{\bm{x}}'(\bm{x}^{k},\bm{z}^{k}; s^k) < 0$ according to the Direction Related Assumption (defined above) and $\beta ^k \ge 0$, above expression can be rewritten as $f(\bm{x}^{k} + \beta ^k s^k,\bm{z}^{k}) < f(\bm{x}^{k},\bm{z}^{k})$. We also have $f(\bm{x}^{k+1},\boldsymbol{z}^{k+1}) \le f(\bm{x}^{k} + \beta ^k s^k,\bm{z}^{k}) < f(\bm{x}^{k},\bm{z}^{k})$ and $f(\bm{x}^{k+1},\boldsymbol{z}^{k+1}) < f(\bm{x}^{k},\bm{z}^{k})$. Also, $f$ is bounded from below, we have $\lim _{k \rightarrow \infty } f(\bm{x}^{k},\bm{z}^{k}) = \bar{f} > -\infty$. Hence, we have
\begin{align} 
\lim _{k \rightarrow \infty } f\left( \bm{x}^{k+1},\boldsymbol{z}^{k+1}\right) - f\left( \bm{x}^{k},\bm{z}^{k}\right) = 0. \nonumber
\end{align}
Furthermore,
\begin{align} 
\lim _{k \rightarrow \infty } f\left( \bm{x}^{k} + \beta ^k s^k,\bm{z}^{k}\right) - f\left( \bm{x}^{k},\bm{z}^{k}\right) = 0. 
\end{align}
For the sake of contradiction, we assume that $\lim _{k \rightarrow \infty } (\bm{x}^{k},\bm{z}^{k}) = (\bar{\bm{x}},\bar{\bm{z}})$ does not satisfy the convergence condition \eqref{convergence_condition}. From the definition of gradient related assumption \cite{bertsekas1997nonlinear}, we have
\begin{align} 
\limsup _{k \rightarrow \infty } f_{\bm{x}}'\left( \bm{x}^{k},\bm{z}^{k} ; s^k\right) < 0. 
\end{align}
Hence, in conclusion, $\lim _{k \rightarrow \infty } \beta ^k = 0$.
From Armijo rule, after a certain iteration $k \ge \bar{k}$, we can define $\left\{ \bar{\beta }^k \right\}$, $\bar{\beta }^k = \beta ^k / \gamma$ for some $\gamma$, where $\bar{\beta }^k \le 1$ and we have
\begin{align} 
\sigma f_{\bm{x}}'\left( \bm{x}^{k},\bm{z}^{k} ; s^k\right) < \frac{f\left( \bm{x}^{k} + \bar{\beta }^k s^k,\bm{z}^{k}\right) - f\left( \bm{x}^{k},\bm{z}^{k}\right) }{\bar{\beta }^k}. 
\end{align}
If we apply the mean value theorem to the right side of the above expression, for some $\widetilde{\beta }^k \in [0,\bar{\beta }^k]$, we have
\begin{align} 
\sigma f_{\bm{x}}'\left( \bm{x}^{k},\bm{z}^{k} ; s^k\right) < f_{\bm{x}}'\left( \bm{x}^{k} +\widetilde{\beta }^k s^k, \bm{z}^{k} ; s^k\right). \label{proof23}
\end{align}
Moreover, $\limsup _{k \rightarrow \infty } f_{\bm{x}}'(\bm{x}^{k},\bm{z}^{k} ; s^k) < 0$, and if we take a limit point $\bar{s}$ of $\left\{ s^k \right\}$ such that $f_{\bm{x}}'(\bar{\bm{x}},\bar{\bm{z}},\bar{s}) < 0$. Also, we have, $\lim _{k \rightarrow \infty , k \in \mathcal {K}} f_{\bm{x}}'(\bm{x}^{k},\bm{z}^{k} ; s^k) = f_{\bm{x}}'(\bar{\bm{x}},\bar{\bm{z}} ; \bar{s})$ and $\lim _{k \rightarrow \infty , k \in \mathcal {K}} f_{\bm{x}}'(\bm{x}^{k} +\widetilde{\beta }^k s^k, \bm{z}^{k} ; s^k) = f_{\bm{x}}'(\bar{\bm{x}},\bar{\bm{z}} ; \bar{s})$. From these two factors, we can infer that $f_{\bm{x}}'(\bm{x},\bm{z};s)$ is continuous. Now, from expression \eqref{proof23}, we have
\begin{align} 
\sigma f_{\bm{x}}'(\bar{\bm{x}},\bar{\bm{z}} ; \bar{s}) \le f_{\bm{x}}'(\bar{\bm{x}},\bar{\bm{z}} ; \bar{s}) \quad \Longrightarrow \quad 0 \le (1-\sigma ) f_{\bm{x}}'(\bar{\bm{x}},\bar{\bm{z}} ; \bar{s}). \nonumber
\end{align}
Since $(1-\sigma ) > 0$, $f_{\bm{x}}'(\bar{\bm{x}},\bar{\bm{z}} ; \bar{s})<0$ which is a contradiction. Therefore, the limit point $(\bar{\bm{x}},\bar{\bm{z}})$ of the sequence  $\left\{ (\bm{x}^{k},\bm{z}^{k}) \right\}$ i.e., $\lim _{k \rightarrow \infty } (\bm{x}^{k},\bm{z}^{k}) = (\bar{\bm{x}},\bar{\bm{z}})$ satisfies the convergence condition, which means algorithm 1 always converges.

\noindent
{\it Part 2: The limit point $(\bar{\bm{x}},\bar{\bm{z}})$ of the sequence  $\left\{ (\bm{x}^{k},\bm{z}^{k}) \right\}$ generated by Algorithm 1 is a global optimal solution of the MICP subproblems.}

\noindent
From \textit{Part 1}, we have that the algorithm converges to the limit point $(\bar{\bm{x}},\bar{\bm{z}})$. In other words, the algorithm produces a solution, $(\bar{\bm{x}},\bar{\bm{z}})$. Here, we establish the global optimality of the solution $(\bar{\bm{x}},\bar{\bm{z}})$. The optimality conditions (KKT conditions) associated with the $(\bar{\bm{x}},\bar{\bm{z}}) \in {{\mathrm{argmin}}}_{\bm{x},\bm{z}} \left\{ L_\gamma (\bm{x},\bm{z},\boldsymbol{\lambda} ) :\bm{\alpha _{\mathrm{}}} \in \bm{\alpha _{\mathrm{}}}^k\right\}$ is given as follows:
\begin{align} 
\Phi _{\bm{x}} :=&\left[ \begin{array}{c} \nabla f(\bar{\bm{x}}) + [\boldsymbol{\lambda} + \gamma (\bar{\bm{x}}-\bar{\bm{z}})]^\top \bf 1 \end{array}\right] ^\top \left[ \begin{array}{c} \bm{x}-\bar{\bm{x}} \end{array}\right] \nonumber\\
&\ge 0 \nonumber .
\end{align}
Note that integer (binary) variables are fixed here. The above optimality condition can also be written as:
\begin{align} 
\min _{\bm{x}} \left\{ 
\Phi _{\bm{x}} \right\} = 0.\nonumber 
\end{align}
The above expression can be re-written in terms of $\check{\varphi }(\boldsymbol{\lambda} + \gamma (\bar{\bm{x}}-\bar{\bm{z}}), \bar{\bm{x}})$ as:
\begin{align} \check{\varphi }(\boldsymbol{\lambda} + \gamma (\bar{\bm{x}}-\bar{\bm{z}}), \bar{\bm{x}})&= f(\bar{\bm{x}}) + \boldsymbol{\lambda} ^\top \bar{\bm{x}} + \gamma \left\| \bar{\bm{x}} - \bar{\bm{z}} \right\| _2^2 \nonumber \\
&= L_\gamma (\bar{\bm{x}},\bar{\bm{z}},\boldsymbol{\lambda} ) + \frac{\gamma }{2} \left\| \bar{\bm{x}} - \bar{\bm{z}} \right\| _2^2.\nonumber \end{align}
We have,
\begin{align} 
\check{\varphi }(\boldsymbol{\lambda} ,\bm{x}^{k}) :=& \min _{\bm{x}} \left\{ f(\bm{x}^{k}) + \nabla _{\bm{x}} f(\bm{x}^{k})^\top (\bm{x}-\bm{x}^{k}) \right . \nonumber\\
&\left . + \boldsymbol{\lambda} ^\top \bm{x} : \bm{x} \in \mathcal{X} \right\} . \nonumber
\end{align}
Note that according to \cite{lubin2013parallelizing}, minimizing linear objective function over mixed-integer convex sets $\mathcal{X}_{\mathrm{p}}$ and $\mathcal{X}_{\mathrm{v}}$ is equivalent to minimizing linear objective function over convex hulls sets, $\text{CH}(\mathcal{X}_{\mathrm{p}})$ and $\text{CH}(\mathcal{X}_{\mathrm{v}})$. Also, we have
\begin{align}
\hat\varphi\left( \bar{\bm{x}},\bar{\bm{z}}, \boldsymbol{\lambda} \right):= &L_\gamma (\bar{\bm{x}},\bar{\bm{z}},\boldsymbol{\lambda} ) + \frac{\gamma }{2} \left\| \bar{\bm{x}} - \bar{\bm{z}} \right\| _2^2.\nonumber
\end{align}
Hence,
\begin{align} \check{\varphi }(\boldsymbol{\lambda} + \gamma (\bar{\bm{x}}-\bar{\bm{z}}), \bar{\bm{x}})=\hat\varphi\left( \bar{\bm{x}},\bar{\bm{z}}, \boldsymbol{\lambda} \right). \label{proof1}
\end{align}
The expression \eqref{proof1} implies that the upper and lower bounds of the Lagrangian function converge as $k \rightarrow \infty$. In other words, Algorithm 1 converges to the global optimal solution of the centralized implementation of MICP subproblems. \cite{conn1991globally}. 

\vspace{-3mm}
\section{Case Study}
This section presents case studies and simulation results. First, the simulation setup is described. Second, the advantages of the proposed framework and the algorithm are illustrated via simulation results. 
\vspace{-3mm}
\subsection{Simulation setup}
\begin{figure}[htbp]
\centering
\includegraphics[scale=.5]{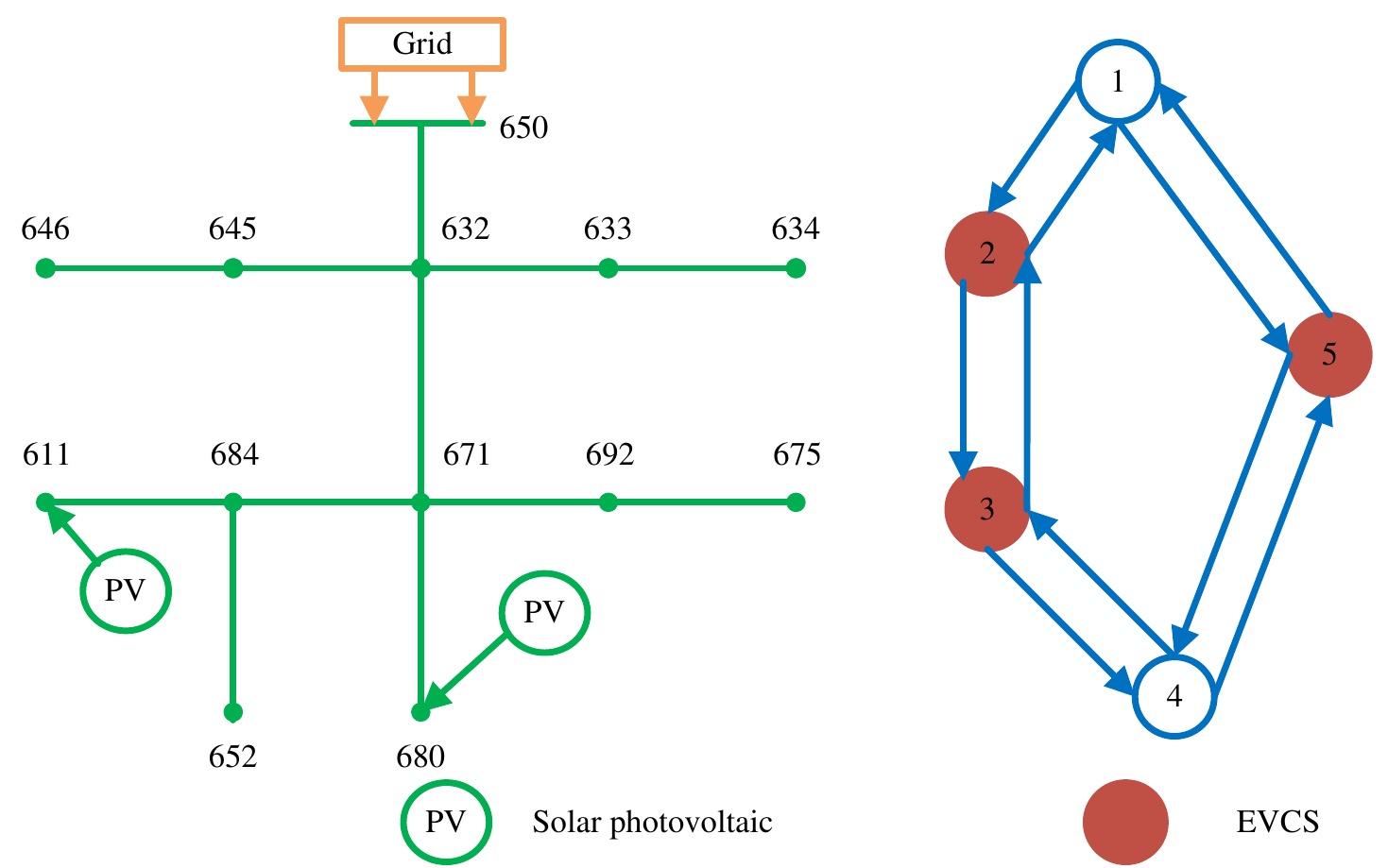}
\caption{Feeder and TN topology (Case 1).}
\label{Case1}
\end{figure}

\begin{figure}[htbp]
\centering
\includegraphics[scale=.5]{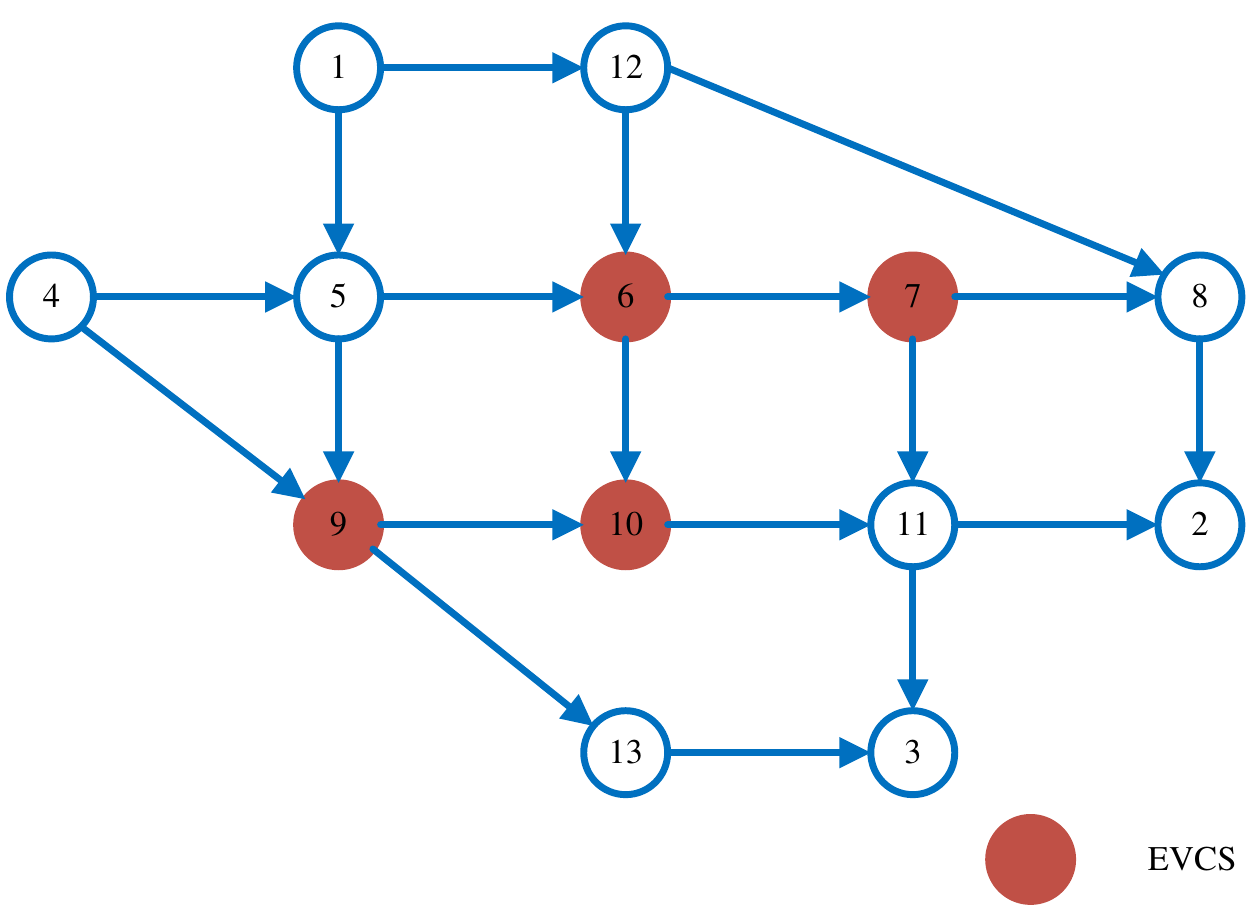}
\caption{TN topology (Case 2).}
\label{Case2}
\end{figure}

\begin{figure}[htbp]
\centering
\includegraphics[scale=.45]{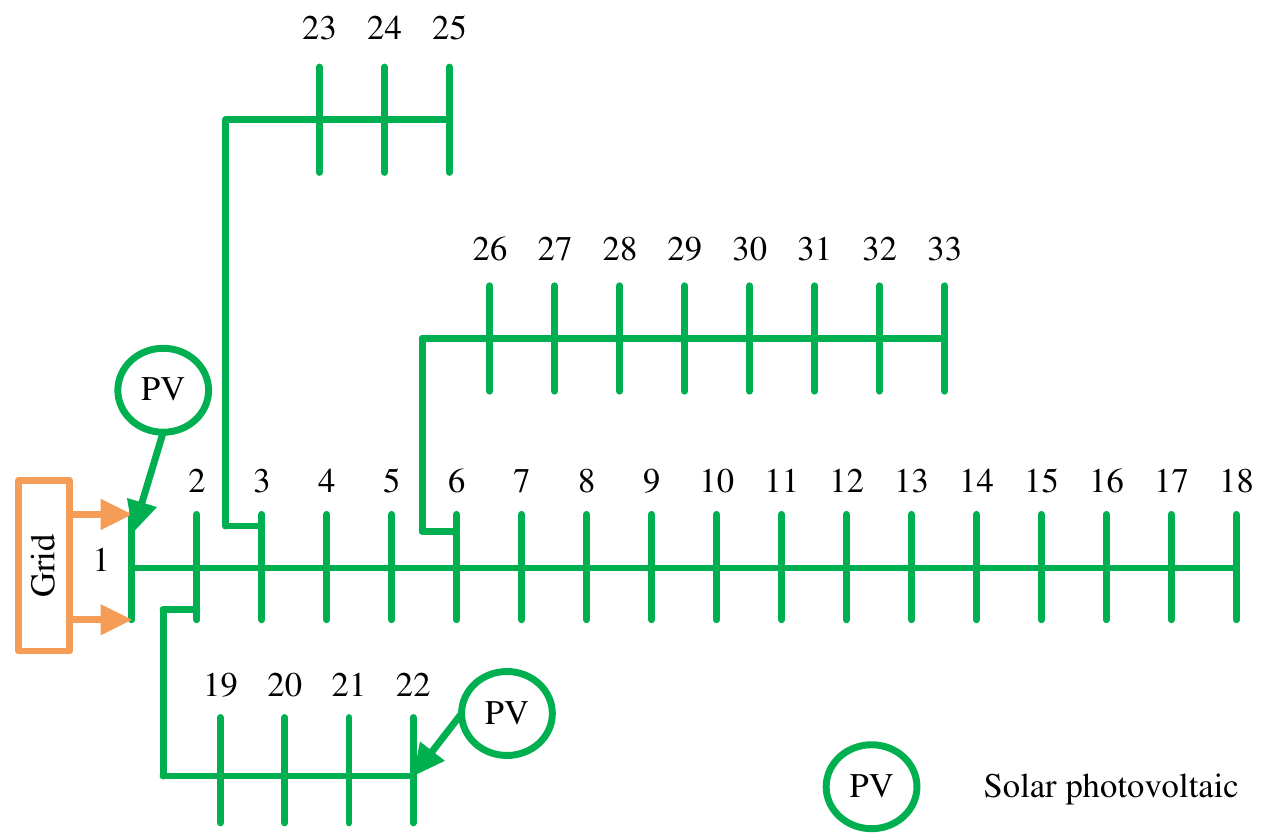}
\caption{Feeder topology (Case 2).}
\label{Case22}
\end{figure}

Generally, the coverage area of TN is much larger than that of  a power distribution feeder \cite{9330803}. Therefore, to make the area of coverage similar, one power distribution feeder supplies one EVCS in a TN in this paper. For Case 1, the three modified IEEE 13-node test feeders \cite{kersting1991radial} represent the PDN, while the 5-node road network represents the TN, as shown in Figure \ref{Case1}. For Case 2, the four modified IEEE 33-node test feeders \cite{kersting1991radial} are used to represent the PDN, while the modified Nguyen-Dupius network is adopted to represent the TN, as shown in Figure \ref{Case2} and \ref{Case22}, respectively. It is worth mentioning that feeders are not coupled with each other, and they are supplied by different buses of the transmission network (grid), hence different grid prices (can be interpreted as LMPs of transmission system). The details of the physical coupling between TN and PDN and the grid prices used are provided in Table \ref{physical_coupling}. In Figures \ref{Case1}, \ref{Case2}, and \ref{Case22}, the network drawn with green color represents PDN while the network drawn with blue color represents TN.
 
\begin{table}[htbp]
\centering
\caption{Information on the physical coupling between TN and PDN and grid prices.}
\label{physical_coupling}
\resizebox{\columnwidth}{!}{%
\begin{tabular}{ccccccc}
\hline \hline
 & \multicolumn{3}{c}{Case 1} & \multicolumn{3}{c}{Case 2} \\ \hline
 & \begin{tabular}[c]{@{}c@{}}TN \\ Node\end{tabular} & \begin{tabular}[c]{@{}c@{}}Feeder \\ Node\end{tabular} & \begin{tabular}[c]{@{}c@{}}Grid Price\\  (\$/MWh)\end{tabular} & \begin{tabular}[c]{@{}c@{}}TN \\ Node\end{tabular} & \begin{tabular}[c]{@{}c@{}}Feeder \\ Node\end{tabular} & \begin{tabular}[c]{@{}c@{}}Grid Price\\  (\$/MWh)\end{tabular} \\ \hline
Feeder 1 & 2 & 633 & 70.47 & 6 & 3 & 70.47 \\ \hline
Feeder 2 & 3 & 650 & 77.52 & 7 & 4 & 77.52 \\ \hline
Feeder 3 & 5 & 680 & 84.57 & 9 & 5 & 84.57 \\ \hline
Feeder 4 &  &  &  & 10 & 6 & 91.62 \\ \hline \hline
\end{tabular}%
}
\end{table}

\begin{table}[htbp]
\centering
\caption{TN Arcs Parameters (Case 1).}
\label{TN_Arcs_1}
\begin{tabular}{ccccc}
\hline \hline
Start & End & \begin{tabular}[c]{@{}c@{}}$t_a^0$\\ (min.)\end{tabular} & \begin{tabular}[c]{@{}c@{}}$c_a$\\ (per min.)\end{tabular} & \begin{tabular}[c]{@{}c@{}}$d_a$\\ (miles)\end{tabular} \\ \hline
1/2 & 2/1 & 120 & 40 & 150 \\ \hline
2/3 & 3/2 & 130 & 40 & 150 \\ \hline
3/4 & 4/3 & 120 & 40 & 150 \\ \hline
1/4 & 4/1 & 195 & 47 & 225 \\ \hline
4/5 & 5/4 & 195 & 47 & 225 \\ \hline \hline
\end{tabular}%
\end{table}

\begin{table}[htbp]
\centering
\caption{Algorithm parameters (Case 1 and 2).}
\label{parameters_used}
\resizebox{\columnwidth}{!}{%
\begin{tabular}{ccccccccc}
\hline \hline
 & $\epsilon/\epsilon_u$ & $\gamma$ & $K$  & $\boldsymbol{z}^0$ &$\boldsymbol{\lambda _{{\mathrm{p}}}}^0 \backslash \boldsymbol{\lambda _{\mathrm{v}}}^0$ &$\check{\varphi}_{\mathrm{p}}^0$ & $\check{\varphi}_{\mathrm{v}}^0$ \\ \hline
Case 1 &  6e-3/1e-1&  4e-6&  300&   $\bm{0}$ &  $\bm{0}$ &  -9999&  -99999  \\ \hline
Case 2 &  6e-3/1e-1&  4e-6&  300&   $\bm{0}$ &  $\bm{0}$ &  -9999&  -99999  \\ \hline \hline
\end{tabular}%
}
\end{table}

The arcs parameters for TN of Case 1 are provided in Table \ref{TN_Arcs_1}, while that of Case 2 are adopted from \cite{9079677}. For Case 1, two O-D pairs considered are 1$\rightarrow$4 and 4$\rightarrow$1 with a traffic (EV) demand of 30 each while, for Case 2, four O-D pairs considered are 1$\rightarrow$2, 1$\rightarrow$3, 4$\rightarrow$2, and 4$\rightarrow$3 with a traffic demand of 60 each. The capacity of solar photovoltaics used in the PDN of both cases is 200 KW. Moreover, the parameters used for the proposed algorithm are provided in Table \ref{parameters_used}.

It should be noted that in the TN model, all feasible paths for each origin-destination pair are used as input. However, in a larger network, the number of feasible paths can become overwhelming. Additionally, not all feasible paths for a given origin-destination pair are actually used by electric vehicles, as many of them are much longer and thus more costly than shorter alternatives. To address this issue, we have narrowed down the set of feasible paths by following a specific rule. First, all paths for each origin-destination pair are generated. Then, any paths that do not contain at least one EVCS node are removed as infeasible. Finally, any feasible paths that are longer than twice the length of the shortest path are also removed.

\subsection{Case-1: IEEE 13-node PDN and 5-node TN}
\begin{figure}[htbp]
\centering
\includegraphics[scale=0.6]{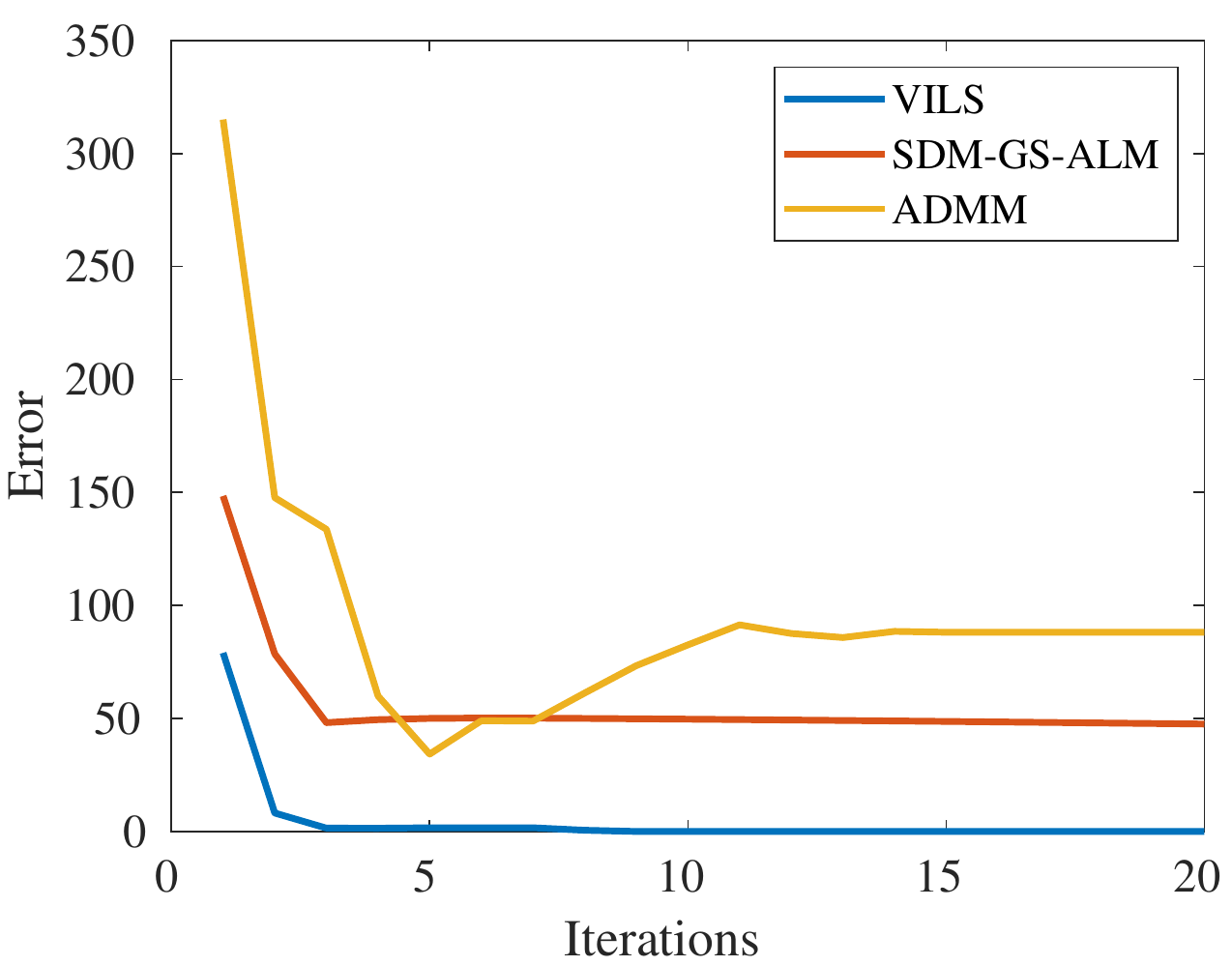}
\caption{Comparison of the convergence error of the proposed VILS algorithm and SDM-GS-ALM algorithm (Case 1).}
\label{ConvergenceError1}
\end{figure}

This section exhibits the results of the coordination of TN and PDN through numerical experiments on a test system, shown in Figure \ref{Case1}. The proposed algorithm is compared with the ADMM \cite{boyd2011distributed} and the SDM-GS-ALM \cite{boland2019parallelizable} (with one inner loop) as presented in Figure \ref{ConvergenceError1}. The figure illustrates that the proposed VILS algorithm outperforms both ADMM and SDM-GS-ALM (with one inner loop) as they failed to converge after 300 main iterations. For better visualization, the convergence error of the ADMM was scaled down by a factor of 10, and the convergence error of only 20 iterations is shown in Figure \ref{ConvergenceError1}. Finally, Table \ref{Case1_results} provides the routing and charging of EVs along with the path flow. For example, for the O-D pair 1-4, 7 EVs are routed on path 1$\rightarrow$\textcircled{2}$\rightarrow$\textcircled{3}$\rightarrow$4 while 23 EVs are routed on path 1$\rightarrow$\textcircled{5}$\rightarrow$4. Note that the circled node number indicates where the EVs are recharged.

\begin{table}[htbp]
\centering
\caption{Charging and routing of EVs (Case 1)}
\label{Case1_results}
\begin{tabular}{cc|cc}
\hline \hline
O-D pair & Path (flow) & O-D pair & Path (flow) \\ \hline
1$\rightarrow$4 & \begin{tabular}[c]{@{}c@{}}1$\rightarrow$\textcircled{2}$\rightarrow$\textcircled{3}$\rightarrow$4 (7)\\ 1$\rightarrow$\textcircled{5}$\rightarrow$4 (23)\end{tabular} & 4$\rightarrow$1 & \begin{tabular}[c]{@{}c@{}}4$\rightarrow$\textcircled{3}$\rightarrow$\textcircled{2}$\rightarrow$1 (7)\\ 4$\rightarrow$\textcircled{5}$\rightarrow$1 (23)\end{tabular} \\ \hline \hline
\end{tabular}%
\end{table}

\vspace{-4mm}
\subsection{Case-2: IEEE 33-Node PDN and Modified Nguyen-Dupius TN}
\begin{figure}[htbp]
\centering
\includegraphics[scale=0.6]{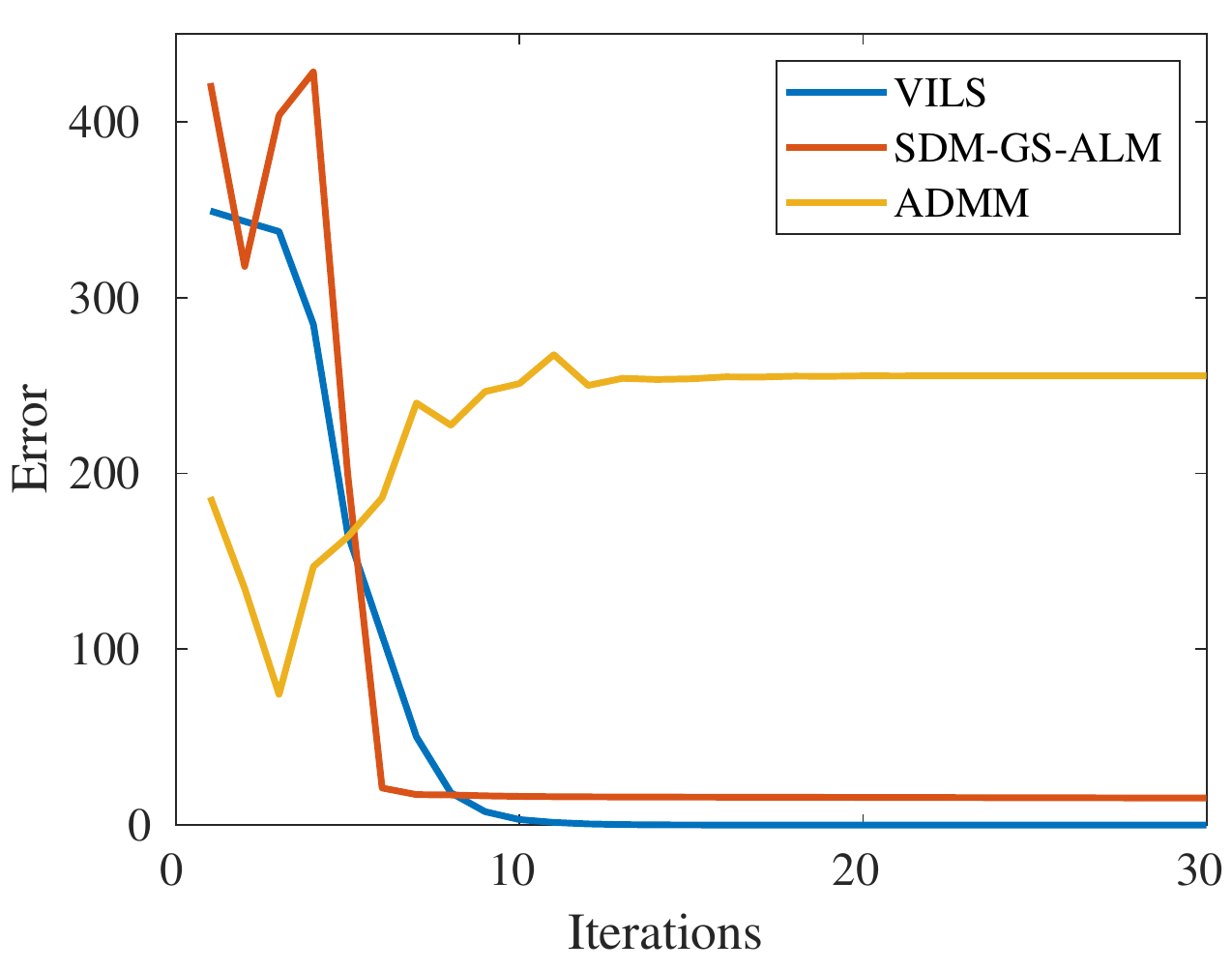}
\caption{Comparison of the convergence error of the proposed VILS algorithm and SDM-GS-ALM algorithm (Case 2).}
\label{ConvergenceError2}
\end{figure}

This section provides the results of the coordination
of TN and PDN through numerical experiments on a bigger
system. The topology of the transportation network and 
the power distribution feeder is shown in Figures \ref{Case2} and \ref{Case22}, respectively. As in Case 1, the proposed algorithm is compared with the ADMM \cite{boyd2011distributed} and the SDM-GS-ALM \cite{boland2019parallelizable} (with one inner loop) as presented in Figure \ref{ConvergenceError2}. The figure illustrates that the proposed VILS algorithm outperforms ADMM and SDM-GS-ALM (with one inner loop) as they failed to converge after 200 main iterations. For better visualization, the convergence error of the ADMM was scaled down by a factor of 100, and the convergence error of only 30 iterations is shown in Figure \ref{ConvergenceError2}. Lastly, Table \ref{Case2_results} provides the routing and charging of EVs along with the path flow.

\begin{table}[htbp]
\centering
\caption{Charging and routing of EVs (Case 2)}
\label{Case2_results}
\begin{tabular}{cc}
\hline \hline
O-D pair & Path (flow) \\ \hline
1$\rightarrow$2 & 1$\rightarrow$5$\rightarrow$6$\rightarrow$\textcircled{7}$\rightarrow$8$\rightarrow$2 (60) \\ \hline
4$\rightarrow$2 & 4$\rightarrow$5$\rightarrow$\textcircled{6}$\rightarrow$\textcircled{7}$\rightarrow$8$\rightarrow$2 (60) \\ \hline
1$\rightarrow$3 & 1$\rightarrow$5$\rightarrow$\textcircled{6}$\rightarrow$\textcircled{10}$\rightarrow$11$\rightarrow$3 (60) \\ \hline
4$\rightarrow$3 & 4$\rightarrow$\textcircled{9}$\rightarrow$\textcircled{10}$\rightarrow$11$\rightarrow$3 (60) \\ \hline \hline
\end{tabular}%
\end{table}
\vspace{-3mm}
\subsection{Computational performance}
In this subsection, the computational performance of the proposed VILS algorithm is compared with the SDM-GS-ALM \cite{boland2019parallelizable}, and the centralized implementation of coordinated d-OPF and OTF, as shown in Table \ref{tab:my-table}. It is worth noting that this paper is not pursuing computational efficiency over the centralized model. However, the proposed algorithm outperforms the existing decentralized algorithms with similar features. For example, in Case 1, the SDM-GS-ALM (with 1 inner loop) failed to converge. However, with 8 inner loops, the SDM-GS-ALM converged in 13 main iterations compared to 10 main iterations of the proposed VILS algorithm, as shown in Table \ref{tab:my-table}. Nonetheless, the SDM-GS-ALM required more than three times more inner loop iterations (104 total inner loop iterations) than the proposed VILS algorithm (28 total inner loop iterations). Similarly, in Case 2, the proposed VILS algorithm outperformed the SDM-GS-ALM algorithm. Finally, it is worth noting that the proposed algorithm converged to the solutions of the centralized implementation of coordinated d-OPF and OTF, as shown in Table \ref{tab:my-table}.

\begin{table*}[htbp]
\centering
\caption{Comparison of computational performance.}
\label{tab:my-table}
\begin{tabular}{cccccccc}
\hline \hline
 & Algorithm & \begin{tabular}[c]{@{}c@{}}Number of inner \\ loop iterations\end{tabular} & \begin{tabular}[c]{@{}c@{}}Total inner \\ loop iterations\end{tabular} & \begin{tabular}[c]{@{}c@{}}Total outer\\ loop iterations\end{tabular} & Converged? & \begin{tabular}[c]{@{}c@{}}Objective \\ value\end{tabular} & \begin{tabular}[c]{@{}c@{}}Computational\\ time\end{tabular} \\ \hline
\multirow{4}{*}{Case 1} & VILS & N/A & 28 & 10 & Yes & \$29941.16 & 101.1 s \\ \cline{2-8} 
 & \multirow{2}{*}{SDM-GS-ALM} & 1 & 300 & 300 & No & N/A & N/A \\ \cline{3-8} 
 &  & 8 & 104 & 13 & Yes & \$29941.11 & 243.1 s \\ \cline{2-8} 
 & Centralized & N/A & N/A & N/A & Yes & \$29941.21 & 11.2 s \\ \hline
\multirow{4}{*}{Case 2} & VILS & N/A & 56 & 20 & Yes & \$ 49549.04 & 200.4 s \\ \cline{2-8} 
 & \multirow{2}{*}{SDM-GS-ALM} & 1 & 200 & 200 & No & N/A & N/A \\ \cline{3-8} 
 &  & 5 & 110 & 22 & Yes & \$ 49549.02 & 342.9 s \\ \cline{2-8} 
 & Centralized & N/A & N/A & N/A & Yes & \$ 49549.13 & 19.2 s \\ \hline \hline
\end{tabular}%
\end{table*}

\vspace{-3mm}
\section{Conclusion}
This paper presents a novel fully decentralized algorithm for the coordination of PDN and TN with EVs. In comparison to existing methods, the main benefits of the proposed algorithm are: 1) unlike existing algorithms like ADMM, it is applicable to MIP problems with convergence and optimality guaranteed; 2) it only requires limited information exchange between PDN and TN operators, which will help preserve the privacy of the two systems and reduce the investment in building communication channels, 3) it is fully decentralized so that all the computations are carried out by PDN operator and TN coordinator only, 4) it is faster than SDM-GS-ALM. The proposed algorithm was tested on two PDN and TN test cases. The test results showed the significance of the proposed framework and algorithm over the existing ones.
\vspace{-4mm}
\bibliographystyle{IEEEtran}

\bibliography{JournalRef.bib}

\end{document}